 \documentclass[final,3p,times]{elsarticle}

\usepackage{amssymb}
\usepackage{amsmath} 
\usepackage{graphicx}
\usepackage{geometry}
\usepackage{textcomp}
\usepackage{color}
\usepackage{hyperref}
\usepackage{lineno}
\usepackage{multicol}
\journal{J.  Quantitative Spectroscopy and Radiative Transfer}

\begin{document}

\begin{frontmatter}



\title{Relation between Raman backscattering from droplets and bulk water: Effect of refractive index dispersion\footnote{\copyright 2017 This manuscript version is made available under the CC-BY-NC-ND 4.0 license \url{http://criativecommons.org/licenses/by-nc-nd/4.0/}}}


\author[taras]{Taras Plakhotnik  \corref{mycorrespondingauthor}}
\cortext[cor_author]{Corresponding author}
\ead{taras@physics.uq.edu.au}

\author[jens]{Jens Reichardt }
\address[taras]{School of Mathematics and Physics, The University of Queensland, St Lucia, QLD 4072, Australia}
\address[jens]{Richard-A\ss mann-Observatorium, Deutscher Wetterdienst, Am Observatorium 12, Lindenberg 15848, Germany}
\begin{abstract}
A theoretical framework is presented that permits investigations of the relation between inelastic backscattering from microparticles and bulk samples of Raman-active materials. It is based on the Lorentz reciprocity theorem and no fundamental restrictions concerning the microparticle shape apply. The approach provides a simple and intuitive explanation for the enhancement of the differential backscattering cross-section in particles in comparison to bulk. The enhancement factor for scattering of water droplets in the diameter range from 0 to 60\,\textmu m (vitally important for the \emph{a priori} measurement of liquid water content of warm clouds with spectroscopic Raman lidars) is about a factor of 1.2-1.6 larger (depending on the size of the sphere) than an earlier study has shown. The numerical calculations are extended to 1000\,\textmu m and demonstrate that dispersion of the refractive index of water becomes an important factor for spheres larger than 100  \textmu m. The physics of the oscillatory phenomena predicted by the simulations is explained. 
\end{abstract}


\end{frontmatter}



\section{Introduction}
The water content of clouds, be it in liquid or frozen form, is one of the key parameters that govern the energy budget of the atmosphere, and thus the weather and by extension the climate of the Earth \cite{Turner_2007,Comstock_2007}. For this reason accurate measurements of cloud water content are of high importance so that microphysical processes in clouds can be studied and eventually understood better, and numerical weather prediction and climate models may be validated. Over the years, remote sensing has become an integral part of such endeavors for the spatial and temporal coverage it provides. Today, both active and passive instruments are monitoring clouds from space and from the ground continuously, and cloud microphysical products are generated routinely from these observations. However, one should take notice of the fact that these products are often the results of retrieval algorithms based on proxy variables and modeling rather than stemming from direct measurements of the parameter itself, which adds another layer of uncertainty. For instance, in the case of ice water content (IWC), common retrieval techniques employ empirical relations between radar reflectivity (e.g., \cite{Hogan_2006,Protat_2007}), or lidar extinction coefficient (e.g., \cite{Heymsfield_2005,Heymsfield_2014}), and IWC derived from ice particles sampled \emph{in situ} during field campaigns. So, ideally, direct measurement methods should be devised to verify the retrieval techniques. Our objective is to determine liquid water content (LWC) and IWC from lidar measurements \emph{a priori} by utilizing the Raman effect.

The water molecule is Raman-active in all three phases of matter, and Raman scattering by water vapor has been exploited successfully for lidar measurements of atmospheric humidity for a long time (as an early example of an operational water vapor Raman lidar, see \cite{Goldsmith_1998}). For experimental and methodological reasons, however, Raman lidar studies of the condensed water phases are much more complicated, and despite dedicated efforts over the last years (see the reviews given in \cite{Sakai_2013,Reichardt_2014}), \emph{a priori} LWC and IWC measurements have been proven elusive.  This is about to change with the advent of spectroscopic water Raman lidars. These instruments allow for the first time direct measurement of the Raman backscatter coefficients of cloud water and ice  \cite{Reichardt_2014}. 

Let $\beta$ be the Raman backscatter coefficient of cloud droplets, then
\begin{equation} \label{JR1}
\mathrm{LWC} = \frac{K  \beta}{\mathrm{d}\sigma_\mathrm{s} /\mathrm{d}\Omega},
\end{equation}
where $K $ is a known instrument-specific constant. One can directly obtain LWC from the measurement of $\beta $ provided that  $\mathrm{d}\sigma_\mathrm{s} /\mathrm{d}\Omega$, the Raman differential backscattering cross-section of a water molecule within a water droplet (subscript 's' stands for sphere) is known. A similar relation applies to IWC, only the numerical values of $K$, $\beta$, and $\mathrm{d}\sigma / \mathrm{d}\Omega$ (being shape dependent)  are different. Note, however, that $\mathrm{d}\sigma_\mathrm{s} / \mathrm{d}\Omega$ is not the same as the cross-section $\mathrm{d}\sigma_\mathrm{b} /\mathrm{d}\Omega$ determined in laboratory experiments using bulk samples  (subscript 'b' for bulk), but differs from it substantially and exhibits a size dependence as previous studies have shown \cite{Veselovskii_2002_2,Volkov_2011}.

Let $\eta_\mathrm{s} $ be the ratio of the molecular cross-section in a droplet to the one in the bulk water sample, henceforth called the enhancement factor:
\begin{equation} \label{JR2}
\eta_\mathrm{s}  = \frac{\mathrm{d}\sigma_\mathrm{s} /\mathrm{d}\Omega}{\mathrm{d}\sigma_\mathrm{b} /\mathrm{d}\Omega},
\end{equation}
then Eq.(\ref{JR1}) can be rewritten as:
\begin{equation} \label{JR3}
\mathrm{LWC} = \frac{K  \beta}{\eta_\mathrm{s}  \,\, \mathrm{d}\sigma_\mathrm{b} /\mathrm{d}\Omega}.
\end{equation}
So in order to obtain LWC \emph{a priori}, we have to determine the Raman differential backscattering cross-section of a water molecule in a macrosample and the magnitude of the size-dependent enhancement factor. In a previous publication, we have obtained $\mathrm{d}\sigma_\mathrm{b} /\mathrm{d}\Omega$ with high accuracy \cite{Plakhotnik_2017}, the subject of the present paper is the investigation of $\eta_\mathrm{s}$. Because the situation is even more complicated for ice due to the enhancement factor being dependent on the shape of the ice particle \cite{Sprynchak_2003,Weigel_2006}, we  focus here mostly on  the liquid phase. The enhancement factor for ice particles will be discussed in a follow-up article.

Incidentally, we point out that a study of the enhancement factor of water droplets was published previously \cite{Veselovskii_2002_2} which, however, was restricted to relatively small size parameters and left some questions unaddressed. Thus our motivation has been threefold: (1) Find a simple and intuitive explanation for the enhancement of the molecular Raman backscattering cross-section in water droplets in comparison to bulk samples. (2) Determine the magnitude of $\eta_\mathrm{s}$. Because any error in $\eta_\mathrm{s}$ directly affects LWC results, this knowledge is crucial. (3) Extend the droplet size range to diameters of drizzle and small rain drops for which a spherical shape may still be assumed, and explore the dependence of $\eta_\mathrm{s}$ on size.

The article is organized as follows. In Section~2, the theory of our model is described in detail. We have followed a new approach and have applied the Lorentz reciprocity theorem to the analysis of Raman scattering by particles. The numerical results are presented and discussed in Section~3. Conclusions are drawn and an outlook is given in Section~4.

\section{Theory}
\label{}

The following theory is basic and is not limited to the case of spherical liquid droplets. To evaluate the value of $\eta$, we use a new approach based on Lorentz reciprocity theorem \cite{landau} which states that  for any volume and its enclosing surface $S$  the following relation between the volume and surface integrals
\begin{equation}
\int[\vec{J}_1\vec{E}_2-\vec{J}_2\vec{E}_1]\mathrm{d}V=\oint_S[\vec{E}_1\times\vec{H}_2-\vec{E}_2\times\vec{H}_1]\mathrm{d}\vec{S}
\end{equation}
 holds  for two sinusoidal current densities $\vec{J}_1$ and $\vec{J}_2$ oscillating at the same frequency  and generating the  electromagnetic fields $\vec{E}_1,\vec{H}_1$ and $\vec{E}_2,\vec{H}_2$. For a particular case of $\vec{J}_1$ and $\vec{J}_2$ being the currents of two point dipoles and the  volume covering the whole space, the surface integral vanishes and the theorem simplifies to 
\begin{equation}\label{eq:reciprocity_dip}
\vec{\mu}\vec{E}^{(d)}=\vec{d}\vec{E}^{(\mu)}
\end{equation}
where $\vec{E}^{(d)}$ is the field created by a point dipole $\vec{d}$ at the location of point dipole $\vec{\mu}$ and $\vec{E}^{(\mu)}$ is the field created by  $\vec{\mu}$ at the location of $\vec{d}$. 

Suppose that the point electrical dipole $\vec{\mu}$ is immersed in a dielectric of an arbitrary shape. The dielectric material occupies volume $V$. Both dipoles oscillate at angular frequency $\omega'$. We assume a large distance between the two dipoles (much larger than the size of $V$ and the wavelength of the wave). Without a loss of generality, we can also assume that $\vec{d}'$ is oriented along $x$-axis of the coordinate system and consider a wave radiated by this dipole propagating in $z$-direction towards $\vec{\mu}$. At a large distance  from $\vec{d}'$, the electromagnetic wave emitted by $\vec{d}'$  can be treated as a plane $x$-polarized wave (this wave is considered plane within $V$).  The electrical field of this (\emph{pumping}) wave reads ${E}_0\exp(k'z-i\omega't)$, where  ${E}_0\propto d'$.

When the \emph{pumping} wave interacts with the dielectric volume, the  internal field (inside the volume)  can be presented as a vector field $\vec{E}_\mathrm{i}^{(x)}(x,y,z,\omega')$, where we drop the time-dependent factor  $\exp(-i\omega't)$ and the superscript indicates that the internal field is calculated for the case of a plain, $x$-polarized  incident wave. Suppose that $(x,y,z)$ is the location of the dipole $\vec{\mu}$ which is induced by $\vec{E} _\mathrm{i}^{(x)}$. In the simplest case of Raman scattering, $\vec{\mu}=\alpha\vec{E} _\mathrm{i}^{(x)}(x,y,z,\omega')$ with $\alpha$ being polarizability but it oscillates with angular frequency $\omega$. The field produced by this dipole is the \emph{scattered} wave and can be obtained from Eq. (\ref{eq:reciprocity_dip}) by considering an auxiliary dipole $\vec{d}$.  Generally, the angular coordinates of this dipole  can be arbitrary, but here we take a practically important case of backscattering when the location of $\vec{d}$ coincides with $\vec{d}'$. For simplicity  it is assumed  that   $|\vec{d}|=|\vec{d}'|$. Vector $\vec{d}$  can be either parallel or perpendicular to $\vec{d}'$. In the case of $\vec{d}\parallel \vec{d}'$, one gets  $\alpha\vec{E} _\mathrm{i}^{(x)}(x,y,z,\omega)\vec{E} _\mathrm{i}^{(x)}(x,y,z,\omega')=dE_x^{(\mu)}$. The projection of the scattered field on $y$-axis can be obtained by considering  $\vec{d}\perp \vec{d}'$ which results in  $\alpha\vec{E} _\mathrm{i}^{(y)}(x,y,z,\omega)\vec{E} _\mathrm{i}^{(x)}(x,y,z,\omega')=dE_y^{(\mu)}$.    

If  there are many \emph{incoherent} induced dipoles homogeneously distributed over the entire volume $V$, then one can get the total power radiated by these dipoles in the direction to  the dipole $\vec{d}$ by integration. The differential $x$-polarized backscattering cross-section per dipole  is the radiant intensity of the scattered wave (proportional to  $|E_x^{(\mu)}|^2$) divided by the intensity (irradiance) of the pumping wave  (proportional to $|E_0|^2$) and similar for the $y$-polarized scattering. Thus, one gets   

\begin{equation}\label{eq:sigma_x}
\frac{\mathrm{d}\sigma^{(x)}_\mathrm{V}}{\mathrm{d}\Omega} =\Upsilon \frac{|\alpha|^2}{|\vec{E}_0|^4}\frac{1}{V}\int_V\left|\vec{E} _\mathrm{i}^{(x)}(x,y,z,\omega)\vec{E} _\mathrm{i}^{(x)}(x,y,z,\omega')\right|^2\mathrm{d}V
\end{equation}
and
\begin{equation}\label{eq:sigma_y}
\frac{\mathrm{d}\sigma^{(y)}_\mathrm{V}}{\mathrm{d}\Omega} =\Upsilon \frac{|\alpha|^2}{|\vec{E}_0|^4}\frac{1}{V}\int_V\left|\vec{E} _\mathrm{i}^{(y)}(x,y,z,\omega)\vec{E} _\mathrm{i}^{(x)}(x,y,z,\omega')\right|^2\mathrm{d}V
\end{equation}
where $\Upsilon$ absorbs all the constant factors such as speed of light in vacuum, concentration of dipoles etc.  This constant also includes a factor dependent on the units,  photon/(s\,sr) or W/sr  used for the radiant intensity. The value of  the total backscattering cross-section (a common case of lidar measurements is integration of scattering over both polarizations) can be obtained as a sum of the two values: 
\begin{equation}\label{eq:sigma_tot}
\frac{\mathrm{d}\sigma_\mathrm{V}}{\mathrm{d}\Omega} =\frac{\mathrm{d}\sigma_\mathrm{V}^{(x)}}{\mathrm{d}\Omega}+\frac{\mathrm{d}\sigma_\mathrm{V}^{(y)}}{\mathrm{d}\Omega}.
\end{equation}
\subsection{Bulk Raman scattering}
\begin{figure}[htbp] 
   \centering
   \includegraphics[width=2in]{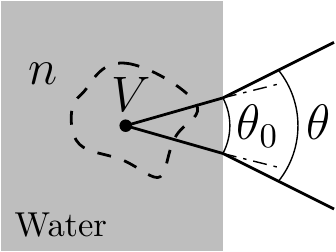} 
   \caption{Bulk experiment. Scattered waves are collected from the molecules occupying volume $V$, the region enclosed by the dashed line. Because for small angles $\theta =n\theta_0$, the differential scattering $\mathrm{d}\sigma/\mathrm{d}\Omega$ is reduced by the factor $n^2$. This factor can be eliminated by immersing the detector of scattering in water but usually the apparent value is simply multiplied by $n^2$ and so it becomes intrinsic to the scattering medium. }
   \label{fig:bulk}
\end{figure}

First, we apply Eqs. (\ref{eq:sigma_x}, \ref{eq:sigma_y}, \ref{eq:sigma_tot})  to the case of bulk scattering. In such a case the dielectric is a large volume (theoretically a half-space) and has a plain interface with air but the scattering is collected from a volume small  in comparison to the size of the bulk sample (see Fig.\,\ref{fig:bulk}). In practice, this volume is defined by the details of the experimental setup. The internal field inside the bulk ${E} _\mathrm{i}^{(x)}(x,y,z,\omega)$ is uniform and in accordance with  Fresnel's formula reads 
\begin{equation}
E _\mathrm{i}^{(x)}(x,y,z,\omega)=\frac{2}{n+1}E_0\exp(ikz),
\end{equation} 
where $k$ is the wave number of light, and similar for the field at frequency $\omega'$.  The $y$-polarized cross-section is zero in this case.  Thus the total bulk differential backscattering cross-section reads
\begin{equation}\label{eq:bulk}
\frac{\mathrm{d}\sigma_\mathrm{b}}{\mathrm{d}\Omega}=\Upsilon|\alpha|^2\frac{16}{(n+1)^2(n'+1)^2}
\end{equation}
where we have allowed for the difference in the refractive index at $\omega$ and $\omega'$. As a matter of fact, it is customary  to take into account the effect of the interface on the scattering and rescale the apparent value of the differential cross-section to its value in the dielectric media \cite{Plakhotnik_2017, Plakhotnik_2013}. 

First, the power transmitted through the interface is reduced at approximately normal incident by the factors $t'=4n'/(n'+1)^2$ and $t=4n/(n+1)^2$ for the pumping wave and for the scattering wave, the expression on the right side of Eq.(\ref{eq:bulk}) should be divided by $tt'$. Second, the solid angle increases by the factor $n^2$ on the interface and therefore the cross-section should be multiplied by $n^2$ when rescaled to the medium (see  Fig.\,\ref{fig:bulk}). This simplifies the  expression for  the differential cross-section to 
\begin{equation}\label{eq:bulk_cor}
\frac{\mathrm{d}\tilde{\sigma}_\mathrm{b}}{\mathrm{d}\Omega} =\frac{n}{n'}\Upsilon |\alpha |^2.
\end{equation}  
The value of $n/n'\approx 0.996$ (for water pumped at 355-nm wavelength) is very close to 1 and this factor will be ignored in the following analysis. But the dispersion will be an important factor when we consider Raman scattering by microparticles. 
\subsection{Relative cross-section of Raman backscattering  by microparticles}

It is practically convenient to compare scattering by microparticles to the scattering by bulk material. For an arbitrary  shaped particle of  volume $V$ one gets from Eqs. (\ref{eq:sigma_x}) and (\ref{eq:bulk_cor}) the enhancement factor for  $x$-polarized scattering 
\begin{equation}\label{eq:vol_rel}
\eta^{(x)}_\mathrm{V} \equiv\frac{\mathrm{d}\sigma^{(x)}_\mathrm{V}/\mathrm{d}\Omega}{\mathrm{d}\tilde{\sigma}_\mathrm{b}/\mathrm{d}\Omega}= \frac{1}{V|\vec{E}_0|^4}\int_V \left|\vec{E} _\mathrm{i}^{(x)}(x,y,z,\omega)\vec{E} _\mathrm{i}^{(x)}(x,y,z,\omega')\right|^2\mathrm{d}V .
\end{equation}
A similar equation for $\eta^{(y)}_\mathrm{V}$ is obtained by replacing $\vec{E} _\mathrm{i}^{(x)}(x,y,z,\omega)$ with $\vec{E} _\mathrm{i}^{(y)}(x,y,z,\omega)$. The total enhancement factor then reads $\eta_\mathrm{V} \equiv \eta^{(x)}_\mathrm{V}+\eta^{(y)}_\mathrm{V}$. The internal fields can be found  numerically, analytically or  using a combination of the two. 

The inhomogeneity of the distribution of the energy density inside $V$ is the main reason for the enhancement factor being larger than 1. The variance of $|E_\mathrm{i}|^2$ is defined by the equation 
\begin{equation}\label{eq:var}
\mathrm{var}(|E_\mathrm{i}|^2)\equiv\frac{1}{V}\int_V|E_\mathrm{i}|^4\mathrm{d}V-\left( \frac{1}{V}\int_V|E_\mathrm{i}|^2\mathrm{d}V\right)^2 .
\end{equation}
We can  use Eqs. (\ref{eq:vol_rel}) and (\ref{eq:var}) to express approximately (ignoring the difference between $\vec{E} _\mathrm{i}^{(x)}(x,y,z,\omega)$ and $\vec{E} _\mathrm{i}^{(x)}(x,y,z,\omega')$) the total enhancement factor as 
\begin{equation}
\eta_\mathrm{V} \approx \frac{\langle|E_\mathrm{i}|^2\rangle^2+\mathrm{var}(|E_\mathrm{i}|^2)}{|E_0|^4}
\end{equation} 
where $\langle\rangle $ stands for the volume averaging. Therefore a more inhomogeneous  distribution of the energy (larger $\mathrm{var}(|E_\mathrm{i}|^2)$) will increase the  relative scattering which is proportional to $\langle|E_\mathrm{i}|^4\rangle$.  

The simplest case of scattering by a microparticle is scattering by a nanosphere with a radius $a$ such that $ka\ll 1$. For such a small sphere, the internal field $E_\mathrm{i}^{(x)}(x,y,z,\omega)$ can be found by solving the corresponding problem in electrostatics and the result reads $E_\mathrm{i}^{(x)}(x,y,z,\omega)= 3/(2+n^2)E_0$.  The $y$-polarized field is zero also in this case. Thus one gets from Eq. (\ref{eq:vol_rel}) 
\begin{equation}\label{eq:nano_rel}
\eta_\mathrm{n} =\frac{81}{(2+n^2)^4}\approx0.40,
\end{equation}
where the numerical value is calculated for water, $n=1.33$. Note that in \cite{Veselovskii_2002_2} this value is close to 0.3 (Fig.\,3b in the cited paper).  Note that Eqs. (24) and (26) in Ref.\cite{Volkov_2011} and Eq.($14'$) in Ref. \cite{Chew_1988} agree with our Eq.(\ref{eq:nano_rel}).

In the following section, we will consider spherical particles large in comparison to the wavelength and will use  Mie theory where the field is expressed in a form of  an infinite series which should be evaluated and integrated numerically. The computations can be accelerated by using spherical coordinates for vectors and space locations because the dependence of the field on the azimuthal angle $\phi$ is very simple:   
\begin{equation}
\vec{E}^{(x)}(r,\theta,\phi,\omega')=\vec{E}_\mathrm{c}^{(x)}(r,\theta,\omega')\cos\phi+\vec{E}_\mathrm{s}^{(x)}(r,\theta,\omega')\sin\phi .
\end{equation}
Moreover, the internal field induced by a plane wave polarized along $y$-axis can be obtained from $\vec{E}^{(x)}(r,\theta,\phi,\omega')$  if $\phi$ is replaced by $\phi+\pi/2$. That is 
\begin{equation}
\vec{E}^{(y)}(r,\theta,\phi,\omega)=\vec{E}_\mathrm{c}^{(x)}(r,\theta,\omega)\sin\phi-\vec{E}_\mathrm{s}^{(x)}(r,\theta,\omega)\cos\phi .
\end{equation}
For briefness, we drop the explicit arguments in the notations of the vector fields and move \emph{prime}  from $\omega'$ to $\vec{E}$. Then due to the mutual orthogonality of $\vec{E}_\mathrm{c}$ and $\vec{E}_\mathrm{s}$
\begin{equation}
|\vec{E}^{(x)}\vec{E}'^{(x)}|^2=|\vec{E}_\mathrm{c}\vec{E}'_\mathrm{c}\cos^2\phi+\vec{E}_\mathrm{s}\vec{E}'_\mathrm{s}\sin^2\phi|^2
\end{equation}
and
\begin{equation}
|\vec{E}^{(x)}\vec{E}'^{(y)}|^2=|\vec{E}_\mathrm{c}\vec{E}'_\mathrm{c}-\vec{E}_\mathrm{s}\vec{E}'_\mathrm{s}|^2\cos^2\phi\sin^2\phi . 
\end{equation}

The integration over $\phi$ can  be done analytically to obtain  
\begin{equation}
\eta^{(x)}_\mathrm{s}=\frac{\pi}{4}\left(3I_1+3I_2+2I_3\right)
\end{equation}
and
\begin{equation}
\eta^{(y)}_\mathrm{s} =\frac{\pi}{4}\left(I_1+I_2-2I_3\right) ,
\end{equation}
where the three double integrals are expressed as follows: 
\begin{equation}\label{eq:I_1}
I_1=\frac{1}{VE_0^4}\int_0^a\int_0^\pi|\vec{E}_\mathrm{c}\vec{E}'_\mathrm{c}|^2r^2\sin\theta  \mathrm{d}\theta \mathrm{d}r
\end{equation}
\begin{equation}\label{eq:I_2}
I_2=\frac{1}{VE_0^4}\int_0^a\int_0^\pi |\vec{E}_\mathrm{s}\vec{E}'_\mathrm{s}|^2r^2\sin\theta  \mathrm{d}\theta \mathrm{d}r
\end{equation}

\begin{equation}\label{eq:I_3}
I_3=\frac{1}{VE_0^4}\int_0^a\int_0^\pi\mathrm{Re}\left[(\vec{E}_\mathrm{c}\vec{E}'_\mathrm{c})(\vec{E}_\mathrm{s}\vec{E}'_\mathrm{s})^*\right]r^2 \sin\theta \mathrm{d}\theta \mathrm{d}r .
\end{equation}
The value of the total relative scattering cross-section (the common case for lidars) can be obtained as a sum of the two values, and the enhancement factor in the case of Raman scattering by a sphere reads: 
\begin{equation}\label{eq:eta_sphere}
\eta_\mathrm{s} =\pi(I_1+I_2).
\end{equation}
\section{Numerical modeling and discussion}

\begin{figure}[htbp] 
   \centering
   \includegraphics[width=12cm, angle =90]{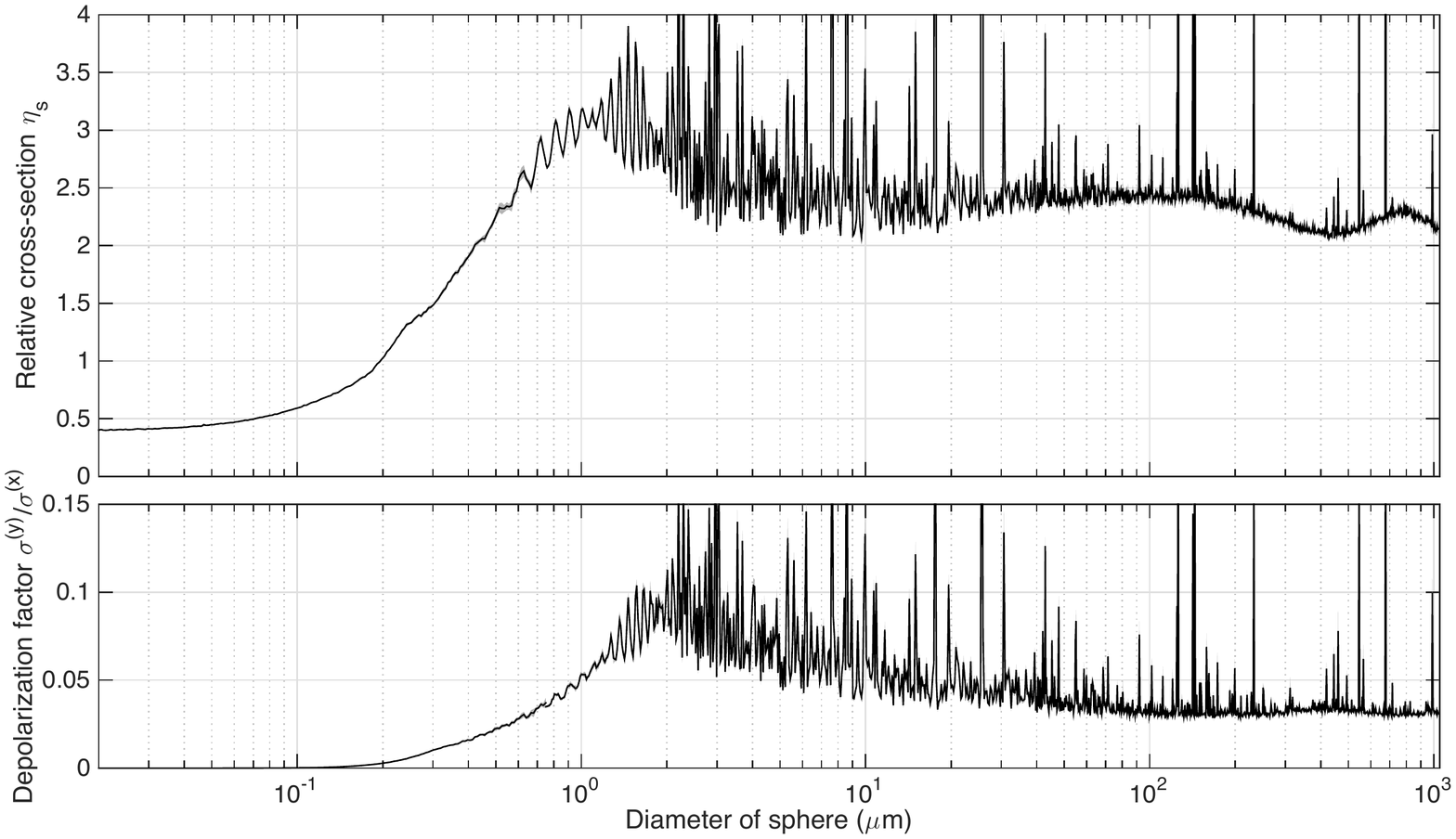} 
   \caption{Relative  Raman differential backscattering cross-section (top) and depolarization factor (bottom) of spheres on a semilogarithmic scale. The wavelengths of pumping and scattered light are 355 nm and  404 nm, respectively. The curves are calculated using correct UV values of the refractive indices ($n_\mathrm{pw}=1.350$ and $n_\mathrm{sw}=1.344$). Accurate calculation of the scattering near the  resonances  (showing up as spikes) has not been attempted (except for a short segment shown in Fig.\,\ref{fig:high_res}).}
\label{fig:raman_rel}
\end{figure}

We have used these equations to calculate Raman scattering of water by spheres of radius $a$ covering the range  from 0 up to 500 \textmu m. The results are shown in Fig.\,\ref{fig:raman_rel}.  Because Raman lidars used to study inelastic scattering by clouds, such as the RAMSES instrument \cite{Reichardt_2014, Reichardt2012}, usually operate at 355 nm, this wavelength has been selected in the computations for the pumping light. The Raman spectrum of liquid water is shifted by 3400 cm$^{-1}$ to a longer wavelength. Refractive indices $n_\mathrm{pw}=1.350$ and   $n_\mathrm{sw}=1.344$ for the pumping and scattered wavelength respectively have been  taken from \cite{refr_index}. The apparently  marginal dispersion of $\Delta n =0.006$ turns out to be an important factor. The electrical field inside the spheres has been obtained  using a standard series expansion in Bessel and spherical harmonic functions \cite{theory_book}.  For the integrations, the electrical field at $4\times 10^4$ points ($16\times 10^4$ points  for spheres larger than 300\,\textmu m) within  the cross-section of the sphere, that is $200 \times 200$  points  ($400 \times 400$)  in the  $(r, \theta)$ space have been used.  The calculations have been done using Matlab code which routinely provides double precision for all numerical values. 

First, we compare Fig.\,\ref{fig:raman_rel} to the results reported by Veselovskii  \cite{Veselovskii_2002_2}, where the size parameter of spheres varies from zero to $\chi \equiv 2\pi a / \lambda=500$ (about 60 \textmu m in diameter for the pumping wavelength of  355 nm). We note that the refractive index used in \cite{Veselovskii_2002_2} is 1.33 for both wavelengths instead of the correct UV values, but the small variation of the refractive index  has a  minor affect on the relative cross-section in this range of $\chi$. For example, the values of $\eta_\mathrm{n}$ obtained with  Eq.(\ref{eq:nano_rel}) are 0.383 and 0.40 for  $n=1.347$ and $n=1.33$, respectively. The value of $\eta_\mathrm{s}$  for the smallest spheres in Fig.\,\ref{fig:raman_rel} is 0.395 (slightly larger than the value calculated with  Eq.(\ref{eq:nano_rel}) but it converges to 0.383 in the limit $a\to0$). The value of $\eta_\mathrm{n}$ reported in \cite{Veselovskii_2002_2}  (see Fig. 3b there) is about a factor of 1.25 smaller than  the theoretical value of 0.40. Figure \,\ref{fig:raman_rel} also shows a peak value of $\eta_\mathrm{s}$ (reached at $\chi\approx 10$ or about 1 \,\textmu m in diameter)   a factor of 1.65 larger than in \cite{Veselovskii_2002_2}. 

\begin{figure}[htbp] 
   \centering
   \includegraphics[width=7cm, angle =90]{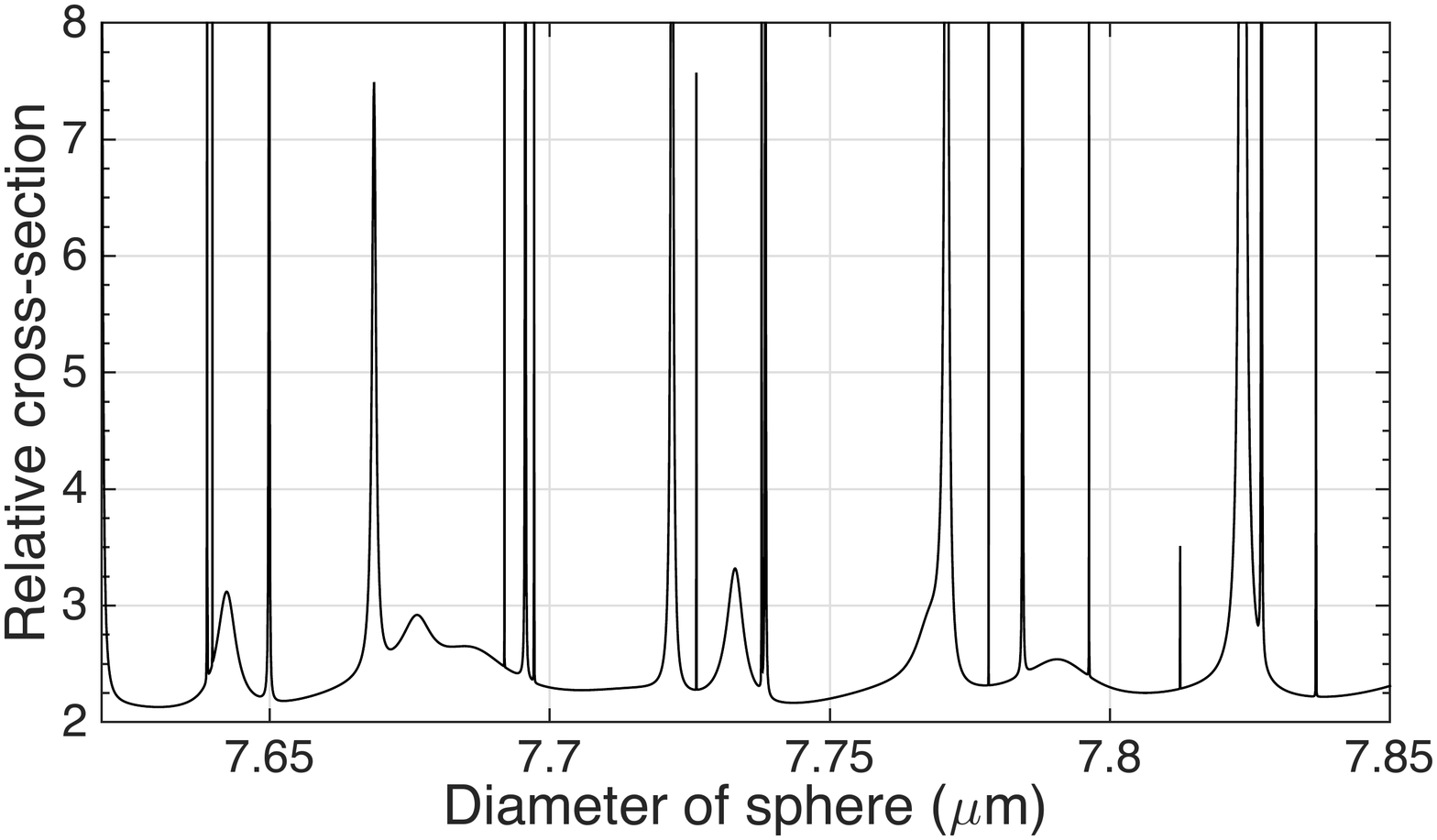} 
   \caption{A short segment of Fig.\,\ref{fig:raman_rel} calculated with a high resolution (the step equals $10^{-6}$ of the sphere diameter). The amplitude and width of the narrow resonances depend on the imaginary part of the refractive index which is assumed to be  $10^{-7}$ in this example. The peak value for the strongest resonance line is  320. }
   \label{fig:high_res}
\end{figure}

 In the range  $30<\chi<70$ (diameters of 3.4 - 7.9\,\textmu m), the previously reported relative backscattering cross-section is about 2.0 in average.  This value includes averaging over resonances. Away from the resonances the value of $\eta_\mathrm{s}$ can be as small as 1.5  (Fig.\,6a in \cite{Veselovskii_2002_2}).  Figure \ref{fig:high_res} shows $\eta_\mathrm{s}$ for a small select range of diameters  (similar to Fig.\,6a in \cite{Veselovskii_2002_2}). Because the contribution of very narrow resonances strongly depends on the  morphology  of the droplets  \cite{MDR}, intrinsic optical losses in water, presence of dust and other impurities, we have tested this dependence by considering three cases. In a theoretical case of zero losses, the average value of $\eta_\mathrm{s}$ is about 3.9. In a more realistic case \cite{water_abs} when the imaginary part of the refraction index $\mathrm{Im}[n]=10^{-8}$, the average value reduces to 2.95. It decreases to 2.85 if $\mathrm{Im}[n]=10^{-7}$. The off-resonance values  are not affected by such a small loss and the minimum value of $\eta_\mathrm{s}$ for the range of diameters shown in Fig.\,\ref{fig:high_res} is  2.13.  Both numbers 2.85 and 2.13 are  a factor  of 1.45 larger than the corresponding values reported by Veselovskii \cite{Veselovskii_2002_2} .   Overall, in the range of diameters covered in \cite{Veselovskii_2002_2} the previosly reported values of $\eta_\mathrm{s}$ are systematically smaller than those of Fig.\,\ref{fig:raman_rel}  but  the two sets of data can not be brought into agreement   by a single scaling factor. The oscillatory behavior  observed for large diameters manifested in Fig.\,\ref{fig:raman_rel} is a novel phenomenon not reported in  earlier publications and will be discussed later in the paper. 

The accuracy of our calculations has been verified in several ways. To assess the limitations of the double precision, a few points  on the curve (100\,\textmu m, 200\,\textmu m and 600\,\textmu m) have been calculated with  quadruple precision (this takes time about a factor of 200  longer  than the double precision calculations) using a multi-precision package \cite{mp_soft} developed by Advanpix LLC. The change of the  calculated value of $\eta_\mathrm{s}$ was less than $10^{-5}$ even at the largest size of the sphere (which requires the largest number of terms in the series expansion of the field). Additionally, the effect of truncation of the infinite series expansion  for the field has been estimated. Increase of the length of the series by 40\% (in comparison to the conventionally used estimate $\chi+4\chi^{1/3}+2$  for the number of required terms) has changed  $\eta_\mathrm{s}$ by about  $10^{-7}$. The main error in the calculations is due to the limited number of the  points used in the final integration step.  Monte Carlo integration technique  has been employed to estimate a 95\% confidence interval.  The  points have been randomly chosen in  the $\theta$-$r$ plane and repeated several times. The estimate of the 95\% confidence interval is  obtained using Student's \emph{t}-distribution with an appropriate number of degrees of freedom (one less than the number of repetitions).

\begin{figure}[htbp] 
   \centering
   \includegraphics[width=12cm, angle =90]{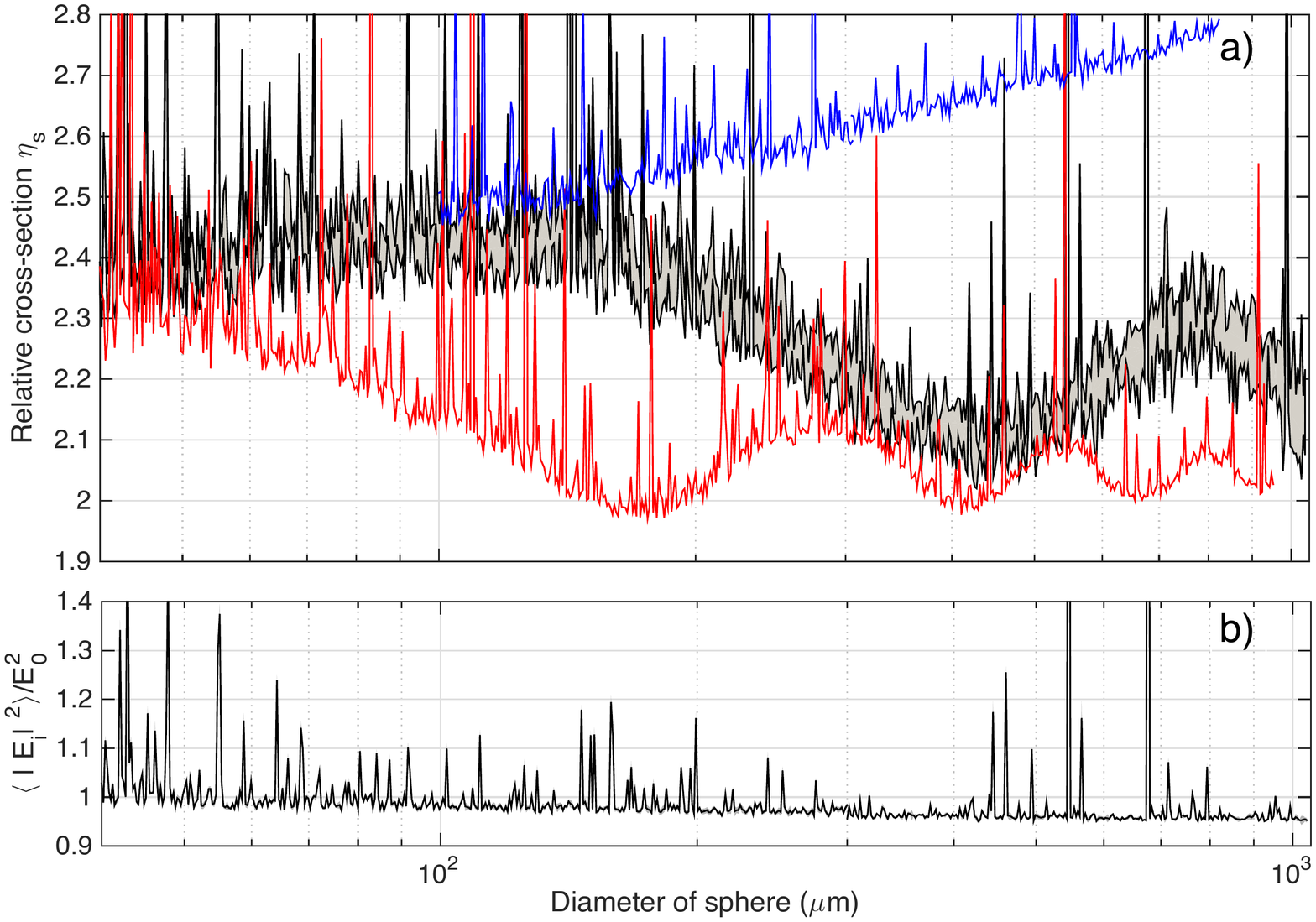} 
   \caption{ a) Relative  Raman differential backscattering cross-section for large diameters  on a semilogarithmic scale. The blue curve shows $\eta_\mathrm{s}$  calculated with a wavelength-independent refractive index of 1.347 (no dispersion), the mean value of the refractive indices at 355 nm and 404 nm. The red curve is obtained for the case of doubled dispersion  ($n_\mathrm{pw}=1.353$ and $n_\mathrm{sw}=1.341$).  Results presented in Fig.\,\ref{fig:raman_rel} are shown as a reference (the grey area marks the 95\% confidence interval). b)  Normalizsed volume average of the electromagnetic field inside the sphere on a semilogarithmic scale.  }
\label{fig:verify}
\end{figure}

The average of $|E_\mathrm{i}|^2$   has been used as another check of the computational accuracy (the total energy of the electromagnetic field inside the sphere equals $nV\langle|E_\mathrm{i}|^2\rangle$). Away from the resonances and  for diameters significantly larger than the wavelength of light, the  geometric optics approximation can be used to show \cite{energy_total_geom} that the volume integral of $|E_\mathrm{i}|^2$ does not depend on the size of the sphere and the wavelength of the pumping light. The volume average reads  
\begin{equation}
\frac{1}{VE^2_0}\int|E_\mathrm{i}|^2\mathrm{dV}=\frac{1}{n^2}\left[(n^3-\left(n^2-1\right)^{3/2}\right] \approx 0.94
\end{equation}
where the numerical value is calculated for water ($n=1.347$, dispersion ignored). This theoretical result agrees with our numerical results which show the value of $\langle|E_\mathrm{i}|^2\rangle/|E_0|^2$ to be close to 0.95 (see Fig.\,\ref{fig:verify}) at large diameters (away from resonances).

\begin{figure}[htbp] 
   \centering
   \includegraphics[width=12cm, angle =-90]{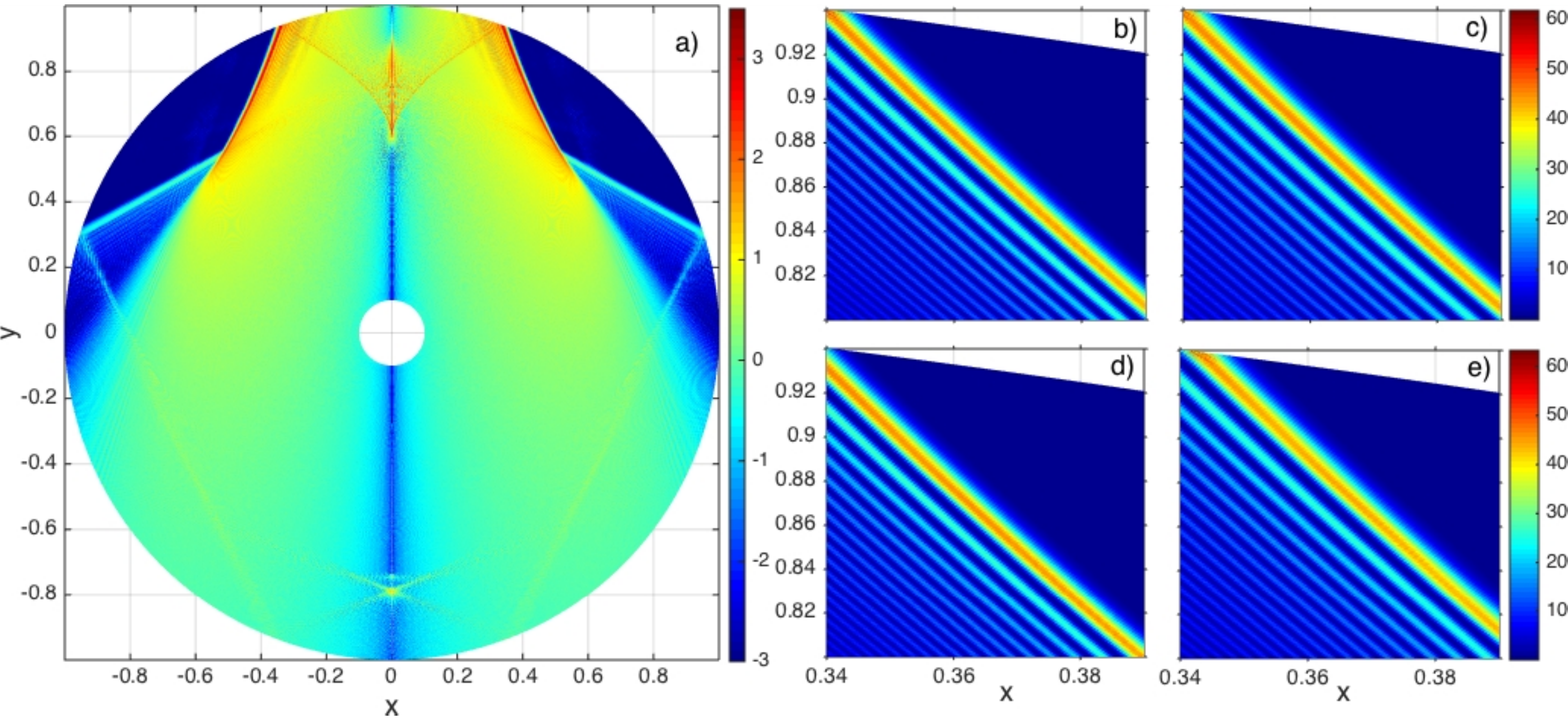} 
   \caption{a) Value of $0.5x\int_0^{2\pi}|\vec{E}\vec{E}'|^2\mathrm{d}\phi$ on a logarithmic scale ($\log_{10}$) across the entire 300-\textmu m sphere. Correct UV refractive indices have been assumed. The value is shown using Cartesian coordinates $x\equiv \pm ra^{-1}\sin\theta$ and  $y\equiv ra^{-1}\cos\theta$. The factor $x$ in front of the integral reduces its value near $x\approx 0$ which correctly reflects the relative insignificance of this region for the volume integral.   For the sake of testing the integration accuracy, these images have been calculated on a $1000\times2000$ grid and the integrals over the  volume resulted in $\eta_\mathrm{s}=2.212$, in agreement with Fig.\,\ref{fig:raman_rel}. Panels b) and c) show $0.5x\int_0^{2\pi}|\vec{E}'|^2\mathrm{d}\phi$ and $ 0.5x\int_0^{2\pi}|\vec{E}|^2\mathrm{d}\phi$ respectively for the strongest features on a linear scale for a 780-\textmu m sphere when $n_\mathrm{pw}=n_\mathrm{sw}=1.347$. Panels d) and e) show the same region of the sphere as in b) and c) but for the case of $n_\mathrm{pw}=1.350$ and $n_\mathrm{sw}=1.344$.  }
   \label{fig:spheres}
   \end{figure}

 Figure \ref{fig:spheres}a illustrates the distribution of $|\vec{E}\vec{E}'|^2$ in a large 300-\textmu m sphere. The distribution is quite inhomogeneous and this results in the enhancement factor being significantly larger than 1. Such distributions for spheres larger than approximately 50 \textmu m  closely resemble each other (with corresponding geometrical scaling) and the results obtained in geometrical optics approximations \cite{energy_dens_geom}, except for the resonances and some features which do not simply scale with the size of the sphere as expected in the geometrical optics approximation. These features critically depend on the wavelength of the pumping/scattered wave. The sphere in Fig.\,\ref{fig:spheres}a has been modelled with much higher spatial resolution than  what was used for calculation of the curves shown in Figs.\,\ref{fig:raman_rel} and \ref{fig:verify},  to verify the accuracy of integration. 

 The oscillatory behavior of $\eta_\mathrm{s}$ at large diameters  and correct UV dispersion of $\Delta n=0.006$ has been investigated in some details to confirm that it is not an artifact but a physical phenomenon. The value of $\eta_\mathrm{s}$ has been calculated  for two hypothetical cases: a two-times larger dispersion ($n_\mathrm{pw}=1.353$, $n_\mathrm{sw}=1.341$) and with a zero dispersion ($n_\mathrm{pw}=n_\mathrm{sw}=1.347$). The curves are shown  in Fig.\,\ref{fig:verify}.   In the case of zero dispersion, the oscillation disappears and  the enhancement factor grows approximately logarithmically with increasing diameter. The three curves overlap (except for the resonances) if the diameter is smaller than 40 \textmu m ($a<20$\,\textmu m). The condition $n_\mathrm{pw}-n_\mathrm{sw}  \approx \lambda/a$ apparently defines a minimal value of $a$ such that the off-resonance cross-sections are noticeably affected by the dispersion.  Larger dispersion   results in earlier deviation of the red curve from the "no dispersion" curve. The first minima in the value of $\eta_\mathrm{s}$ on the red curve  is reached at about 175 \textmu m which is followed by a maximum around  290\,\textmu m.  These values for the black curve are about 430\,\textmu m and 780\,\textmu m respectively. Both numbers are approximately 2.6 times smaller for the larger dispersion. The amplitude of the oscillations of the red curve clearly decays with increasing diameter and the enhancement factor converges to approximately 2.05.  This suggests that the oscillations will decay also for the case $\Delta n=0.006$ (this decay is less obvious in Fig.\,\ref{fig:verify} due to the insufficiently long range of diameters). Finally, note that Monte Carlo  integration (which employs a randomized integration grid) eliminates a possibility of an accidental coincidence of the grid nodes with the antinodes of the electric field (such a coincidence would artificially increase the value of the integral).
 
 To explain the discovered oscillations of  $\eta_\mathrm{s}$ and the reason why the dispersion of water plays such an important role, we focus on the narrow and strongest features in the distributions of the field presented with high resolution in Figs.\,\ref{fig:spheres}b-e. Figures \ref{fig:spheres}b and \ref{fig:spheres}c   show that in the absence of dispersion the positions of the diagonal lines are almost identical for the pumping and scattered fields. The difference (about 10\%) in the two wavelengths just slightly affects  the spacing between these lines. In the case of dispersion, the positions of the lines are different for the pumping and the scattered fields due to different refraction  at the interface between water and air (the refraction plays a critical role when the incident field enters the sphere). The mismatch in the locations of the lines for the two fields reduces the value of $|\vec{E}\vec{E}'|^2$ and hence the value of  $\eta_\mathrm{s}$. But $\eta_\mathrm{s}$ partially recovers if lines $1'$, $2'$, etc of $|\vec{E}'|^2$  (the numbering starts from the strongest line in Fig.\,\ref{fig:spheres}d) correspondingly overlap with lines 2, 3, etc of $|\vec{E}|^2$ (see Fig.\,\ref{fig:spheres}e). As demonstrated by the figure, such a resonance is achieved for the size of the sphere of about 780\,\textmu m. This is the diameter when the value of $\eta_\mathrm{s}$ reaches its first maximum in Fig.\,\ref{fig:verify}a. The first minimum is reached by $\eta_\mathrm{s}$ at  430\,\textmu m. This is the size when the position of  line $1'$  in the distribution $|\vec{E}'|^2$  sits between lines 1 and 2 of the distribution $|\vec{E}|^2$. This size is a bit larger than half of the 780\,\textmu m because lines $1'$ and 1 are the strongest and therefore line $1'$ should be positioned closer to line 2 (not in the middle between lines 1 and 2) to achieve a minimal overlap between $|\vec{E}'|^2$ and $|\vec{E}|^2$.      
  
It may look surprising that the relative backscattering cross-section of a sphere  does not converge to the bulk value with increasing  diameter. This is because we consider only a situation when the distance from the sphere to the point of detection (location of $\vec{d}$) is  much larger than the sphere diameter and therefore the contribution of different points of the sphere to the total scattering is not affected by the collecting optics. Therefore the right hand side of Eq. (\ref{eq:eta_sphere}) does not converge to 1 in the limit $a\to0$. The conventional  "bulk measurements" deal with the situation of a plane interface between water and air when the water and air take a half-space each but the scattering is collected only from a small finite size volume. Therefore when the size of the sphere increases, the integration volume should be decreased to a smaller and smaller  fraction of the sphere  for a proper transformation to bulk.

\section{Conclusion}
We have applied  Lorentz reciprocity theorem to the analysis of Raman backscattering by particles. This approach provides a simple and intuitive explanation for the enhancement of the backscattering cross-section in particles in comparison to bulk samples (theoretically considered as objects occupying a half space). The enhancement factor is related to the variance of the energy density within the particle volume.  This theorem also links the standard Mie theory of elastic scattering to  Raman  scattering, and numerical calculations of relative differential Raman backscattering cross-section have been  carried out for  spherical particles up to 1000-\textmu m diameters. These calculations are in qualitative, but not in quantitative, agreement with previously published results, the values  of the relative cross-section reported in this paper are about a factor of 1.2--1.6 larger (depending on the size of the sphere). We have also discovered that the small dispersion of the refractive index of water has a significant effect on Raman scattering by spheres larger than  100 \textmu m. The observed phenomenon systematically  depends on the factor  $\Delta na/\lambda $.  The oscillations are explained by considering resonance phenomena between narrow and wavelength dependent features in the distributions of the electrical field  at pumping and scattered wavelengths.  

The basic theory developed in this article is applicable to small particles of any shape as long as the internal fields can be determined numerically or analytically. If one studies microphysical cloud  properties with lidars, assumption of a spherical shape for the microparticles is a good choice for several reasons: It is a realistic model for cloud and drizzle droplets as well as drops in light precipitation; Mie theory can be used for the computations; and spatial orientation of the particles with respect to the exciting light field is irrelevant which makes the calculations relatively fast. Obviously, the spherical particle model is only sufficient for warm clouds. Below the frost point, the fraction of aspherical particles increases with decreasing temperatures. So in order to measure IWC \emph{a priori}, one needs to employ a different  model for microparticles (see review \cite{Kahnert_shape_rev}) and different numerical methods such as T-matrix etc \cite{Kahnert_numerical_rev} to compute enhancement factor of cold clouds.  

\section{Acknowledgements}
This research has been supported in part by ARC (Australian Research Council) Grant DP0771676 and in part by Deutsche Wetterdienst.






\begin{thebibliography}{10}
\expandafter\ifx\csname url\endcsname\relax
  \def\url#1{\texttt{#1}}\fi
\expandafter\ifx\csname urlprefix\endcsname\relax\def\urlprefix{URL }\fi
\expandafter\ifx\csname href\endcsname\relax
  \def\href#1#2{#2} \def\path#1{#1}\fi

\bibitem{Turner_2007}
D.~D. Turner, A.~M. Vogelmann, R.~T. Austin, J.~C. Barnard, K.~Cady-Pereira,
  J.~C. Chiu, S.~A. Clough, C.~Flynn, M.~M. Khaiyer, J.~Liljegren, K.~Johnson,
  B.~Lin, C.~Long, A.~Marshak, S.~Y. Matrosov, S.~A. McFarlane, M.~Miller,
  Q.~Min, F.~Minnis, W.~O'Hirok, Z.~Wang, W.~Wiscombe, Thin liquid water clouds
  - {T}heir importance and our challenge, Bull. Amer. Meteor. Soc. 88 (2007)
  177--190.
\newblock \href {http://dx.doi.org/10.1175/BAMS-88-2-177}
  {\path{doi:10.1175/BAMS-88-2-177}}.

\bibitem{Comstock_2007}
J.~M. Comstock, R.~d'Entremont, D.~DeSlover, G.~G. Mace, S.~Y. Matrosov, S.~A.
  McFarlane, P.~Minnis, D.~Mitchell, K.~Sassen, M.~D. Shupe, D.~D. Turner,
  Z.~Wang, An intercomparison of microphysical retrieval algorithms for
  upper-tropospheric ice clouds, Bull. Amer. Meteor. Soc. 88 (2007) 191--204.
\newblock \href {http://dx.doi.org/10.1175/BAMS-88-2-191}
  {\path{doi:10.1175/BAMS-88-2-191}}.

\bibitem{Hogan_2006}
R.~J. Hogan, M.~P. Mittermaier, A.~J. Illingworth, The retrieval of ice water
  content from radar reflectivity factor and temperature and its use in
  evaluating a mesoscale model, J. Appl. Meteor. Climatol. 45 (2006) 301--317.

\bibitem{Protat_2007}
A.~Protat, J.~Delano\"{e}, D.~Bouniol, A.~J. Heymsfield, A.~Bansemer, P.~Brown,
  Evaluation of ice water content retrievals from cloud radar reflectivity and
  temperature using a large airborne in situ microphysical database, J. Appl.
  Meteor. Climatol. 46 (2007) 557--572.

\bibitem{Heymsfield_2005}
A.~J. Heymsfield, D.~Winker, G.-J. van Zadelhof, Extinction-ice water
  content-effective radius algorithms for {CALIPSO}, Geophys. Res. Lett. 32
  (2005) L10807.
\newblock \href {http://dx.doi.org/10.1029/2005GL022742}
  {\path{doi:10.1029/2005GL022742}}.

\bibitem{Heymsfield_2014}
A.~Heymsfield, D.~Winker, M.~Avery, M.~Vaughan, G.~Diskin, M.~Deng, V.~Mitev,
  R.~Matthey, Relationships between ice water content and volume extinction
  coefficient from in situ observations for temperatures from 0$^{\circ}$ to
  -86$^{\circ}${C}: {I}mplications for spaceborne lidar retrievals, J. Appl.
  Meteor. Climatol. 53 (2014) 479--505.

\bibitem{Goldsmith_1998}
J.~E.~M. Goldsmith, F.~H. Blair, S.~E. Bisson, D.~D. Turner, Turn-key {R}aman
  lidar for profiling atmospheric water vapor, clouds, and aerosols, Appl. Opt.
  37 (1998) 4979--4990.

\bibitem{Sakai_2013}
T.~Sakai, D.~N. Whiteman, F.~Russo, D.~D. Turner, I.~Veselovskii, S.~H. Melfi,
  T.~Nagai, Y.~Mano, Liquid water cloud measurements using the {R}aman lidar
  technique: current understanding and future research needs, J. Atmos. Ocean.
  Technol. 30 (2013) 1337--1353.

\bibitem{Reichardt_2014}
J.~Reichardt, Cloud and aerosol spectroscopy with {R}aman lidar, J. Atmos.
  Ocean. Technol. 39 (2014) 1946--1963.
\newblock \href {http://dx.doi.org/10.1175/JTECH-D-13-00188.1}
  {\path{doi:10.1175/JTECH-D-13-00188.1}}.

\bibitem{Veselovskii_2002_2}
I.~Veselovskii, V.~Griaznov, A.~Kolgotin, D.~N. Whiteman, Angle- and
  size-dependent characteristics of incoherent raman and fluorescent scattering
  by microspheres. 2. {N}umerical simulation, Appl. Opt. 41 (2002) 5783--5791.

\bibitem{Volkov_2011}
S.~N. Volkov, I.~V. Samokhvalov, D.~Kim, Raman and fluorescent scattering
  matrix of spherical microparticles, Appl. Opt. 50~(21) (2011) 4054--4062.

\bibitem{Plakhotnik_2017}
T.~Plakhotnik, J.~Reichardt, Accurate absolute measurements of the {R}aman
  backscattering differential cross-section of water and ice and its dependence
  on the temperature and excitation wavelength, J. Quant. Spectrosc. Radiat.
  Transf. 194 (2017) 58--64.
\newblock \href {http://dx.doi.org/10.1016/j.jqsrt.2017.03.023}
  {\path{doi:10.1016/j.jqsrt.2017.03.023}}.

\bibitem{Sprynchak_2003}
V.~Sprynchak, C.~Esen, G.~Schweiger, Enhancement of {R}aman scattering by
  deformation of microparticles, Opt. Lett. 28 (2003) 221--223.

\bibitem{Weigel_2006}
T.~Weigel, J.~Schulte, G.~Schweiger, Inelastic scattering by particles of
  arbitrary shape, J. Opt. Soc. Am. A 23 (2006) 2797--2802.

\bibitem{landau}
L.~D. Landau, E.~M. Lifshitz, Electrodynamics of Continuous Media,
  Addison-Wesley, Reading, MA, 1960 \S89.

\bibitem{Plakhotnik_2013}
A.~Bray, R.~Chapman, T.~Plakhotnik, Accurate measurements of the {R}aman
  scattering coefficient and the depolarization ratio in liquid water, Appl.
  Opt. 52 (2013) 2503--2510.
\newblock \href {http://dx.doi.org/10.1364/AO.52.002503}
  {\path{doi:10.1364/AO.52.002503}}.

\bibitem{Chew_1988}
H.~Chew, Total fluorescent scattering cross sections, Phys. Rev. A 37 (1988)
  4107--4110.

\bibitem{Reichardt2012}
J.~Reichardt, U.~Wandinger, V.~Klein, I.~Mattis, B.~Hilber, R.~Begbie,
  {RAMSES}: {G}erman {M}eteorological {S}ervice autonomous {R}aman lidar for
  water vapor, temperature, aerosol, and cloud measurements, Appl. Opt. 51
  (2012) 8111--8131.

\bibitem{refr_index}
A.~H. Harvey, J.~S. Gallagher, J.~M. H.~L. Sengers, Revised formulation for
  refractive index of water and steam as a function of wavelength, temperature
  and density, J. Phys. Chem. Ref. Data 27~(4) (1998) 761--774.

\bibitem{theory_book}
C.~F. Bohren, D.~R. Huffman, Absorption and scattering of light by small
  particles, Wiley-VCH, Weinheim, 2004.

\bibitem{MDR}
J.~M. Dlugach, M.~I. Mishchenko, Effects of nonsphericity on the behavior of
  {L}orenz--{M}ie resonances in scattering characteristics of liquid-cloud
  droplets, J. Quant. Spectrosc. Radiat. Transf. 146 (2014) 227--234.
\newblock \href {http://dx.doi.org/10.1016/j.jqsrt.2014.01.004}
  {\path{doi:10.1016/j.jqsrt.2014.01.004}}.

\bibitem{water_abs}
G.~M. Hale, M.~R. Querry, Optical constants of water in the 200-nm to
  200-$\mu$m wavelength region, Appl. Opt. 12~(3) (1973) 555--563.

\bibitem{mp_soft}
\href{http://www.advanpix.com/}{Multiprecision computing toolbox for matlab},
  Advanpix LLC. (2017).
\newline\urlprefix\url{http://www.advanpix.com/}

\bibitem{energy_total_geom}
H.~M. Lai, P.~T. Leung, K.~L. Poon, K.~Young, Characterization of the internal
  energy density in {M}ie scattering, Opt. Soc. Am. A 8 (1991) 1553--1558.

\bibitem{energy_dens_geom}
D.~Q. Chowdhury, P.~W. Barber, S.~C. Hill, Energy-density distribution inside
  large not absorbing spheres by using {M}ie theory and geometrical optics,
  Appl. Opt. 31 (1992) 3518--3523.

\bibitem{Kahnert_shape_rev}
M.~Kahnert, T.~Nousiainen, H.~Lindqvist, Review: {M}odel particles in
  atmospheric optics, J. Quant. Spectrosc. Radiat. Transf. 146 (2014) 41--58.
\newblock \href {http://dx.doi.org/10.1016/j.jqsrt.2014.02.014}
  {\path{doi:10.1016/j.jqsrt.2014.02.014}}.

\bibitem{Kahnert_numerical_rev}
M.~Kahnert, Numerical solutions of the macroscopic {M}axwell equations for
  scattering by non-spherical particles: {A} tutorial review, J. Quant.
  Spectrosc. Radiat. Transf. 178 (2016) 22--37.
\newblock \href {http://dx.doi.org/10.1016/j.jqsrt.2015.10.029}
  {\path{doi:10.1016/j.jqsrt.2015.10.029}}.

\end{thebibliography}




\end{document}